\documentclass[11pt]{article}
\usepackage{graphicx}

\def\half{\mbox{$\frac{1}{2}$}}
\def\d{\mbox{\rm d}}

\def\dddot#1{\mathinner{\buildrel\vbox{\kern5pt\hbox{...}}\over{#1}}}
\def\ddddot#1{\mathinner{\buildrel\vbox{\kern5pt\hbox{....}}\over{#1}}}

\title{From Lagrangian to Quantum Mechanics with  Symmetries}

\author{MC Nucci}

\date{Dipartimento di Matematica e Informatica,\\[0.2cm] Universit\`a degli Studi di
Perugia  \& INFN Sezione di Perugia, \\[0.2cm] 06123 Perugia, Italy}

\begin{document}
\maketitle
\begin{abstract}
 We present an old and regretfully forgotten method by Jacobi which
allows one to find many Lagrangians of simple classical models and also of
nonconservative systems. We underline that the knowledge of Lie symmetries
generates Jacobi last multipliers and each of the latter yields a Lagrangian.
Then it is shown that Noether's theorem can identify among those Lagrangians
the physical Lagrangian(s) that will successfully lead to quantization. The
preservation of the Noether symmetries as Lie symmetries of the corresponding
Schr\"odinger equation is the key that takes classical mechanics into quantum
mechanics.
 Some examples are presented.

\end{abstract}

\strut\hfill



\section{Introduction}
The inverse problem of calculus of variation has attracted a lot of interest
since in the second half of the XVIII century Euler \cite{Euler1744} and then
Lagrange \cite{Lagrange1760} introduced the direct problem, namely the idea of
linking the solution of a differential equation to the maximum/minimum of a
functional, the celebrated problem of the brachistochrone being indeed the most
famous classical example. It will take hundred of pages to cite all the papers
and books that have been published since up to date. Most authors mark the
birthdate of the inverse problem with  the 1887-papers by either Helmholtz
\cite{Helm1887}  or  Volterra \cite{Volterra1887}. Some other especially among
the Russian speaking researchers pushes the date slightly back to the
1886-paper by Sonin \cite{Sonin1886}. Very few recognize the seminal work by
Jacobi, namely his 1845-paper \cite{Jacobi1845} and his 1842-1843 Dynamics
Lectures published posthumously in 1884 \cite{JacobiVD}, available in English
since 2009 \cite{VDEng}, where he links his last multiplier to the Lagrangian
for any even-order ordinary differential equation (ODE). Actually both Volterra
and Sonin recognize the contribution of Jacobi last multiplier in their papers,
Sonin more explicitly than Volterra since he showed that
his own method involves the Jacobi last multiplier (p.10 in \cite{Sonin1886}).\\

The method of Jacobi last multiplier was enhanced when Lie determined the link
with his symmetries \cite{Lie1874}, a link very easy to implement that allows
to derive many multipliers and therefore Lagrangians.

It is known that a single second-order ODE admits different many Lagrangians
\cite{Whittaker}, but so far there is not a method that can discern the
physical Lagrangian among them, although some were proposed, e.g.
\cite{Manko81}. The same is true for systems of second-order ODEs that admit
more that  one Lagrangian. It has been shown that some systems of second-order
ODEs do not admit a Lagrangian \cite{Douglas41} although  in \cite{Douglas} the
Lagrangian of some of those systems were determined by following Bateman's
statement \cite{Bateman31}, namely {\em finding a set of equations equal in
number to a given set, compatible with it and derivable from a variational
principle} without recourse to any additional set of equations.

 We propose that the physical Lagrangian should be the one that admits the highest
possible number of Noether symmetries \cite{LGQM}, \cite{gallipoli10},
\cite{schgf}. It was proven in \cite{gonzalez88} that the maximal dimension of
the Lie symmetry algebra of a system of $n$ ODEs of second order is $n^2+4n+3$,
and the highest number of corresponding Noether symmetries is $(n^2+3n+6)/2$.
In particular for a single second-order ODE the highest number of Noether
symmetries is five \cite{LeachNoether}.

Consequently we conjecture that the passage from a classical system  to its
quantum analogue should preserve exactly those Noether symmetries, namely the
Noether symmetries of the physical Lagrangian shall become the Lie symmetries
of the corresponding Schr\"odinger equation  \cite{LGQM}, \cite{gallipoli10},
\cite{schgf}.

In this paper after recalling the method of Jacobi last multiplier and its link
to Lie symmetries, we present the simple example of the one-dimensional free
particle: ten inequivalent\footnote{Namely they do not differ by a total
derivative.} Lagrangians are presented, their Noether symmetries  identify  two
Lagrangians that admit the highest number of Noether symmetry: one independent
and one dependent on time. Then the corresponding Schr\"odinger equations are
obtained. Also the case of the nonlinear second-order Riccati equation is
illustrated and its corresponding Schr\"odinger equation obtained.


\section{Jacobi last multiplier}
The method of the Jacobi Last Multiplier \cite{Jacobi44a}, \cite{Jacobi44b},
\cite{Jacobi1845}, \cite{JacobiVD}
 provides a means to determine all the solutions of the partial
differential equation
\begin {equation}
\mathcal{A}f = \sum_{i = 1} ^n a_i(x_1,\dots,x_n)\frac {\partial f} {\partial
x_i} = 0 \label {2.1}
\end {equation}
or its equivalent associated Lagrange system
\begin {equation}
\frac {\d x_1} {a_1} = \frac {\d x_2} {a_2} = \ldots = \frac {\d x_n}
{a_n}.\label {2.2}
\end {equation}
In fact, if one knows the Jacobi Last Multiplier and all but one of the
solutions, then the last solution can be obtained by a quadrature. The Jacobi
Last Multiplier $M$ is given by
\begin {equation}
\frac {\partial (f,\omega_1,\omega_2,\ldots,\omega_{n- 1})} {\partial
(x_1,x_2,\ldots,x_n)}
 = M\mathcal{A}f, \label {2.3}
\end {equation}
where
\begin {equation}
\frac {\partial (f,\omega_1,\omega_2,\ldots,\omega_{n- 1})} {\partial
(x_1,x_2,\ldots,x_n)} = \mbox {\rm det}\left [
\begin {array} {ccc}
\displaystyle {\frac {\partial f} {\partial x_1}} &\cdots &\displaystyle {\frac {\partial f} {\partial x_n}}\\
\displaystyle {\frac {\partial\omega_1} {\partial x_1}} & &\displaystyle {\frac {\partial\omega_1} {\partial x_n}}\\
\vdots & &\vdots\\
\displaystyle {\frac {\partial\omega_{n- 1}} {\partial x_1}} &\cdots
&\displaystyle {\frac {\partial\omega_{n- 1}} {\partial x_n}}
\end {array}\right] = 0 \label {2.4}
\end {equation}
and $\omega_1,\ldots,\omega_{n- 1} $ are $n- 1 $  solutions of (\ref {2.1}) or,
equivalently, first integrals of (\ref {2.2}) independent of each other. This
means that  $M$ is a function of the variables $(x_1,\ldots,x_n)$ and  depends
on the chosen $n-1$ solutions, in the sense that it varies as they vary. The
essential properties of the Jacobi Last Multiplier are:
\begin{description}
\item{ (a)} If one selects a different set of $n-1$ independent
solutions $\eta_1,\ldots,\eta_{n-1}$ of equation (\ref {2.1}), then the
corresponding
 Last Multiplier $N$ is linked to $M$ by the relationship:
$$
N=M\frac{\partial(\eta_1,\ldots,\eta_{n-1})}{\partial(\omega_1,
\ldots,\omega_{n-1})}.
$$
\item{ (b)} Given a non-singular transformation of variables
$$
\tau:\quad(x_1,x_2,\ldots,x_n)\longrightarrow(x'_1,x'_2,\ldots,x'_n),
$$
\noindent then the Last Multiplier $M'$ of  $\mathcal{A'}F=0$ is given by:
$$
M'=M\frac{\partial(x_1,x_2,\ldots,x_n)}{\partial(x'_1,x'_2,\ldots,x'_n)},
$$
where $M$ obviously comes from the $n-1$ solutions of $\mathcal{A}F=0$ which
correspond to those chosen for $\mathcal{A'}F=0$ through the inverse
transformation $\tau^{-1}$.
\item{ (c) } One can prove that each multiplier $M$ is a solution
of the following
 linear partial differential equation: \begin {equation}
\sum_{i = 1} ^n \frac {\partial (Ma_i)} {\partial x_i} = 0,\label {2.5}\end
{equation} or equivalently: \begin {equation} \frac{{\rm d}}{{\rm d} t}(\log
M)+\sum_{i = 1} ^n \frac {\partial a_i} {\partial x_i}=0;\end{equation}
\noindent viceversa every solution $M$ of this equation is a Jacobi Last
Multiplier.
\item{ (d) } If one
knows two Jacobi Last Multipliers $M_1$ and $M_2$ of equation (\ref {2.1}),
then their ratio is a solution $\omega$ of (\ref {2.1}), or, equivalently,  a
first integral of (\ref {2.2}). Naturally the ratio may be quite trivial,
namely a constant. Viceversa the product of a multiplier $M_1$ times any
solution $\omega$ yields another last multiplier
$M_2=M_1\omega$.\end{description}
  Since the existence of a solution/first integral
is consequent upon the existence of symmetry, an alternate formulation in terms
of symmetries was provided by Lie \cite{Lie1874}, \cite {Lie12}. A clear
treatment of the formulation in terms of solutions/first integrals  and
symmetries
 is given by Bianchi \cite {Bianchi18}. If we know
$n- 1 $ symmetries of (\ref {2.1})/(\ref {2.2}), say
\begin {equation}
\Gamma_i = \sum_{j=1}^{n}\xi_{ij}(x_1,\dots,x_n)\partial_{x_j},\quad i = 1,n-
1, \label {2.6}
\end {equation}
Jacobi's last multiplier is given by $M =\Delta ^ {- 1} $, provided that
$\Delta\not = 0 $, where
\begin {equation}
\Delta = \mbox {\rm det}\left [
\begin {array} {ccc}
a_1 &\cdots & a_n\\
\xi_{1,1} & &\xi_{1,n}\\
\vdots & &\vdots\\
\xi_{n- 1,1}&\cdots &\xi_{n- 1,n}
\end {array}\right]. \label {2.8}
\end {equation}
There is an obvious corollary to the results of Jacobi mentioned above. In the
case that there exists a constant multiplier, the determinant is a first
integral.  This result is potentially very useful in the search for first
integrals of systems of ordinary differential equations.  In particular, if
each component of the vector field of the equation of motion is missing the
variable associated with that component, i.e., $\partial a_i/\partial x_i = 0
$, the last multiplier is a constant, and any other Jacobi Last Multiplier is a
first integral.

Another property of the Jacobi Last Multiplier is  its (almost forgotten)
relationship with the Lagrangian, $L=L(t,q,\dot q)$, for any second-order
equation
\begin{equation}
\ddot q=F(t,q,\dot q) \label{geno2}
\end{equation}
i.e. \cite{JacobiVD} (Lecture 10)\footnote{{\em Jacobi's Lectures on Dynamics}
 are finally available in English \cite{VDEng}.}, \cite{Whittaker}
\begin{equation}
M=\frac{\partial^2 L}{\partial \dot q^2} \label{relMLo2}
\end{equation}
where $M=M(t,q,\dot q)$ satisfies the following equation
\begin{equation} \frac{{\rm d}}{{\rm d} t}(\log M)+\frac{\partial F}{\partial
\dot q} =0.\label{Meq}
\end{equation}
Then equation (\ref{geno2}) becomes the Euler-Lagrange equation:
\begin{equation}
-\frac{{\rm d}}{{\rm d} t}\left(\frac{\partial L}{\partial \dot
q}\right)+\frac{\partial L}{\partial q}=0. \label{ELo2}
\end{equation}
The proof is based on taking the derivative of (\ref{ELo2}) with respect to
$\dot q$
 and showing that this yields (\ref{Meq}).
 If one knows a Jacobi last multiplier, then $L$ can be
easily obtained by a double integration, i.e.:
\begin{equation}
L=\int\left (\int M\, {\rm d} \dot q\right)\, {\rm d} \dot q+f_1(t,q)\dot
q+f_2(t,q), \label{lagrint}
\end{equation}
where $f_1$ and $f_2$ are functions of $t$ and $q$ which have to satisfy a
single partial differential equation related to (\ref{geno2}) \cite{laggal}. As
it was shown in \cite{laggal}, $f_1, f_2$ are related to the gauge function
$g=g(t,q)$. In fact, we may assume
\begin{eqnarray}
f_1&=&  \frac{\partial g}{\partial q}\nonumber\\
f_2&=& \frac{\partial g}{\partial t} +f_3(t,q) \label{gf1f2o2}
\end{eqnarray}
where $f_3$ has to satisfy the mentioned partial differential equation
 and $g$ is obviously arbitrary. We remark the importance
 of the gauge function in order
to apply Noether's theorem \cite{Noether} correctly. Therefore we do not
annihilate
 the gauge function.

 In \cite{laggal} it was shown that if one knows several (at least
two) Lie symmetries of the second-order differential equation (\ref{geno2}),
i.e.
\begin {equation}
\Gamma_j =V_j(t,q)\partial_t+G_j(t,q)\partial_q, \quad j = 1,r,
 \label {gensym}
\end {equation}
 then many Jacobi Last Multipliers could be
derived by means of (\ref{2.8}), i.e.
\begin {equation}
{\displaystyle{\frac{1}{M_{nm}}}}=\Delta_{nm} = \mbox {\rm det}\left [
\begin {array} {ccc}
1 &\dot q & F(t,q,\dot q)\\[0.2cm]
V_n &G_n &{\displaystyle{\frac{\d G_n}{\d t} -\dot q\frac{\d V_n}{\d t}}} \\[0.2cm]
V_m &G_m &{\displaystyle{\frac{\d G_m}{\d t} -\dot q\frac{\d V_m}{\d t}}}\\
\end {array}\right],\label {Mnm}
\end {equation}
with $(n,m=1,r)$, and therefore many Lagrangians  can be obtained by means of
(\ref{lagrint}).

\section{Lagrangians for the free particle}
It is well-known since Lie's seminal work \cite{Lie12} that the equation of a
free particle
\begin{equation}\ddot q=0 \label{freepeq}\end{equation}
admits an eight-dimensional Lie symmetry algebra\footnote{Also reported in
\cite{GonGon83}.}, $sl(3,I\!\!R)$, generated by the following operators:
\begin{eqnarray}
X_1 =q t \partial_t+q^2\partial_q,\quad\quad X_2 = q \partial_t,\quad\quad X_3
= t^2 \partial_t+q t \partial_q,\quad\quad
X_4 = q \partial_q,\nonumber\\[0.2cm]
X_5 = t\partial_t,\quad\quad X_6 =  \partial_t,\quad\quad X_7
=t\partial_q,\quad\quad X_8 = \partial_q. \label{gen}
\end{eqnarray} An obvious Jacobi Last multiplier (JLM) of (\ref{freepeq})
 is a constant since
$\dot q$ does not appear in its right-hand side. This also implies that any JLM
is a first integral of (\ref{freepeq}). We can found this trivial JLM and many
others by calculating the inverse of the determinant of the matrix (\ref{Mnm})
for all the possible combinations of two different operators in (\ref{gen}). It
results that ten different  JLM  and consequently as many Lagrangians, by means
of (\ref{lagrint}), can be obtained\footnote{Since $M_{nm}=-M_{mn}$ we have
arbitrarily chosen the sign as we wished.}, i.e.:
\begin{eqnarray}
M_{13}=-\frac{1}{(t \dot q-q)^3} &\Rightarrow& L_{13}=- \frac{1}{2 t^2 (t \dot
q-q)}
+\frac{{\rm d}g}{{\rm d}t}(t,q)\nonumber\\[0.2cm]
 M_{15}=-\frac{1}{\dot q(t \dot q-q)^2} &\Rightarrow& L_{15}=\frac{\dot
q}{q^2}\left(\log(t\dot
q-q)-\log(\dot q)\right) +\frac{{\rm d}g}{{\rm d}t}(t,q)\nonumber\\[0.2cm]
M_{16}=\frac{1}{\dot q^2(t \dot q-q)} &\Rightarrow& L_{16}= \left(\frac{t \dot
q}{q^2}-\frac{1}{q}\right) \left(\log(\dot q)-\log(t\dot q-q)\right)
+\frac{{\rm d}g}{{\rm d}t}(t,q)
\nonumber\\[0.2cm]
M_{17}=-\frac{1}{(t \dot q-q)^2} &\Rightarrow& L_{17}= -\frac{1}{t^2}\log(t\dot
q-q)+\frac{{\rm d}g}{{\rm d}t}(t,q)\nonumber\\[0.2cm]
M_{18}=\frac{1}{\dot q(t \dot q-q)} &\Rightarrow& L_{18}= -\frac{\dot
q}{q}\log(\dot q) -\left(\frac{1}{t}-\frac{\dot q}{q}\right)\log(t\dot
q-q)+\frac{1}{t}(1+\log(q)) +\frac{{\rm d}g}{{\rm d}t}(t,q)\nonumber\\[0.2cm]
M_{26}=-\frac{1}{\dot q^3} &\Rightarrow& L_{26}= -\frac{1}{2\dot q}+\frac{{\rm
d}g}{{\rm d}t}(t,q) \nonumber\\[0.2cm]
M_{28}=\frac{1}{\dot q^2} &\Rightarrow& L_{28}= -\log(\dot q)+\frac{{\rm
d}g}{{\rm d}t}(t,q)
\nonumber\\[0.2cm]
M_{38}=\frac{1}{t \dot q-q} &\Rightarrow& L_{38}=\left(\frac{\dot
q}{t}-\frac{q}{t^2}\right)
\left(\log(t\dot q-q)-1\right)+\frac{{\rm d}g}{{\rm d}t}(t,q)\nonumber\\[0.2cm]
M_{48}=-\frac{1}{\dot q} &\Rightarrow& L_{48}= \dot q (1-\log(\dot
q))+\frac{{\rm d}g}{{\rm d}t}(t,q)
\nonumber\\[0.2cm]
M_{87}=1 &\Rightarrow& L_{87}= \half \dot q^2+\frac{{\rm d}g}{{\rm d}t}(t,q)
\end{eqnarray}
The ten Lagrangians are NOT linked by the gauge function $g=g(t,q)$.  We note
that seven of the matrices (\ref{Mnm}) have determinant equal to zero, i.e.:
 \begin{equation}
\Delta_{12}, \quad \quad\Delta_{14}, \quad \quad\Delta_{24}, \quad\quad
\Delta_{35}, \quad \quad \Delta_{37}, \quad \quad\Delta_{57}, \quad
\quad\Delta_{68}. \label{det0}
\end{equation}
This is in agreement with the application of Jacobi's method to the linear
harmonic oscillator \cite{laggal} that yielded fourteen different Lagrangians
and three matrices having determinant equal to zero.\\
Applying Noether's theorem yields that both $L_{13}$ and $L_{87}$ admit five
Noether point symmetries, the maximum possible \cite{LeachNoether} in the case
of equation (\ref{freepeq}). The main difference between the two Lagrangians is
that one Lagrangian only is independent on time and is the traditional
Lagrangian, namely the kinetic energy $L_{87}$. Moreover the two Lagrangians
admit different Noether point symmetries. In fact Lagrangian $L_{13}$ admits
the following five Noether symmetries and corresponding first integrals of
equation (\ref{freepeq})
\begin{eqnarray}
X_1&\Longrightarrow&Int_1=- \frac{\dot q}{q-t\dot q},  \nonumber \\
X_2&\Longrightarrow&Int_2=\frac{\dot q^2}{2(q-t\dot q)^2}, \nonumber \\
X_3&\Longrightarrow&Int_3=- \frac{1}{q-t\dot q},\nonumber \\
X_4-X_5&\Longrightarrow&Int_4=- \frac{\dot q}{(q-t\dot q)^2},\nonumber \\
X_7&\Longrightarrow&Int_7=-\frac{1}{2(q-t\dot q)^2},
\end{eqnarray}
while Lagrangian $L_{87}$ admits the following five Noether symmetries and
corresponding first integrals of equation (\ref{freepeq})
\begin{eqnarray}
X_3&\Longrightarrow&In_3=- \half (q-t\dot q)^2 ,  \nonumber \\[0.2cm]
X_4+2X_5&\Longrightarrow&In_4=-\dot q(q-t\dot q), \nonumber \\[0.2cm]
X_6&\Longrightarrow&In_6=\half \dot q^2,\nonumber \\[0.2cm]
X_7&\Longrightarrow&In_7=q-t\dot q,\nonumber \\[0.2cm]
X_8&\Longrightarrow&In_8=-\dot q.
\end{eqnarray}
\section{Schr\"odinger equations for the free particle}
The  Schr\"odinger equation for the free particle is
\begin{equation}
2i\psi_t+\psi_{xx}=0. \label{sch}
\end{equation}
We show that this equation can be obtained by considering a generic linear
parabolic equation
\begin{equation}2i\psi_t+f_1(x)\psi_{xx}+f_2(x)\psi_x+f_3(x)\psi=0 \label{sch1}\end{equation}
with $f_k, (k=1,3)$ functions of $x$ to be determined in such a way that
equation (\ref{sch1}) admits the following five Lie symmetries\footnote{We have
identified $q$ with $x$.}
\begin{eqnarray}
X_3&\Rightarrow&\Omega_1 =t^2 \partial_t+xt \partial_x+
\omega_1\partial_{\psi}, \nonumber\\[0.2cm]
X_4+2X_5&\Rightarrow&\Omega_2 = 2t \partial_t+x\partial_x+
\omega_2\partial_{\psi}, \nonumber \\[0.2cm]
X_6&\Rightarrow&\Omega_3 =\partial_t+\omega_3\partial_{\psi},\nonumber \\[0.2cm]
X_7&\Rightarrow&\Omega_4 =t\partial_x+\omega_4\partial_{\psi},\nonumber \\[0.2cm]
X_8&\Rightarrow&\Omega_5 =\partial_t+\omega_5\partial_{\psi}.
\end{eqnarray}
where $\omega_i=\omega_i(t,x,\psi), (i=1,5)$ are functions of $t,x,\psi$ that
have to be determined.  Equation (\ref{sch1}) also admits the following two
symmetries
\begin{equation}
\Omega_6=\psi\partial_{\psi}, \quad\quad\quad
\Omega_{\alpha}=\alpha(t,x)\partial_{\psi} \label{twosym}
\end{equation}
with $\alpha$ any solution of equation (\ref{sch1}) itself, since any linear
partial differential equation possesses these  two symmetries.

 Using the
interactive REDUCE programs \cite{man}, we obtain that
\begin{equation}
f_1=1,\quad f_{2}=f_{3}=0,
\end{equation}
and
\begin{equation}
\omega_1=\half (ix^2 - t)\psi,\quad  \omega_2=\omega_3=\omega_5=0, \quad
 \omega_4=ix \psi .
 \end{equation}
Therefore the Noether symmetries admitted by the Lagrangian $L_{87}$, namely
the physical Lagrangian for the free particle, yield the right Schr\"odinger's
equation (\ref{sch}) and thus the correct quantization procedure is achieved.

We now show that a quantum-correct Schr\"odinger equation can be derived even
from the five Noether symmetries admitted by the time-dependent Lagrangian
$L_{13}$. We consider a generic linear partial differential equation
\begin{equation}
f_{11}(t,x)\psi_{tt}+f_{12}(t,x)\psi_{tx}+f_{22}(t,x)\psi_{xx}+f_{1}(t,x)\psi_{t}
+f_2(t,x)\psi_{x}+f_0(t,x)\psi=0\label{sch2}\end{equation} with $f_{rs}, f_r,
(r,s=1,2), f_0$ functions of $t,x$ to be determined in such a way that equation
(\ref{sch2}) admits the following five Lie symmetries\footnote{We have
identified $q$ with $x$.}
\begin{eqnarray}
X_1&\Rightarrow&W_1 =xt \partial_t+x^2\partial_x+
w_1\partial_{\psi}, \nonumber\\[0.2cm]
X_2&\Rightarrow&W_2 = x \partial_t+
w_2\partial_{\psi}, \nonumber \\[0.2cm]
X_3&\Rightarrow&W_3 =t^2 \partial_t+xt \partial_x+w_3\partial_{\psi},\nonumber \\[0.2cm]
X_4-X_5&\Rightarrow&W_4 =-t\partial_t+x\partial_x+w_4\partial_{\psi},\nonumber \\[0.2cm]
X_7&\Rightarrow&W_5 =t\partial_x+w_5\partial_{\psi}. \label{symsch2}
\end{eqnarray}
where $w_i=w_i(t,x,\psi), (i=1,5)$ are functions of $t,x,\psi$ that have to be
determined.  Equation (\ref{sch2}) also admits the two symmetries
\begin{equation}
W_6=\psi\partial_{\psi}, \quad\quad\quad W_{\beta}=\beta(t,x)\partial_{\psi}
\label{twosymb}
\end{equation}
with $\beta$ any solution of equation (\ref{sch2}) itself.

Using the interactive REDUCE programs \cite{man}, we obtain that equation
(\ref{sch2}) becomes
\begin{equation}
4t^2 \psi_{tt}+ 8t x \psi_{tx}+4x^2\psi_{xx}+ 12t \psi_t + 12 x \psi_x+ 3\psi
=0 \label{sch20}
\end{equation}
with
\begin{equation}
W_1=-\half x\psi,\quad  W_2=W_4=W_5=0, \quad
 W_3=-\half t\psi .
 \end{equation}
Equation (\ref{sch20}) is parabolic  and therefore can be put into its normal
form since its characteristic coordinates are
\begin{equation}\xi=\frac{x}{t},\quad x=x,\quad \psi=\phi(\xi,x).\label{carcoo}\end{equation} Thus
equation (\ref{sch20}) transforms into
\begin{equation}
4 x^2\phi_{xx}+12x \phi_x+3\phi=0 \end{equation} with solution:
\begin{equation}\phi(\xi,x)=\alpha_1(\xi)x^{-1/2}+\alpha_2(\xi)x^{-3/2},\end{equation}
where $\alpha_1, \alpha_2$ are arbitrary functions of $\xi$. This is obviously
in agreement with the quantization of the free particle.

We remark that there is some freedom when imposing equation (\ref{sch2}) to
admit the Lie symmetries (\ref{symsch2}) but in any possible cases a parabolic
equation with characteristic coordinates (\ref{carcoo}) is always obtained. For
example another possibility is \begin{equation}4t^4\psi_{tt}+ 8t^2x
\psi_{tx}+4t^2x^2\psi_{xx}+4t^2(3t+x)\psi_t + 4tx(3t + x)\psi_x
 +( 3t^2 + 4tx + x^2)\psi =0\label{sch2b}\end{equation}
with
\begin{eqnarray}
W_1=-\half\left(1+\frac{x}{t}\right)x\psi,\quad
W_2=\frac{x^2}{2t^2}\log(t)\psi, \quad W_3=-\half (t+x)\psi, \nonumber \\
W_4=\frac{x}{t}\log(t)\psi, \quad W_5=-\half\log(t)\psi .
 \end{eqnarray}
The normal form of equation (\ref{sch2b}) is
\begin{equation}4 x^2\phi_{xx}+(12+4\xi)x
\phi_x+(3+4\xi+\xi^2)\phi=0\end{equation}
 with solution:
 \begin{equation}\phi(\xi,x)= \alpha_1(\xi)x^{-3/2-\xi/2}+\alpha_2(\xi)x^{-1/2-\xi/2},\end{equation}
 also in agreement with the quantization of the free particle.

\section{Schr\"odinger equation for the second-order Riccati equation}
In \cite{nuctam_3} it was shown that the linearizable second-order Riccati
equation, a member of the Riccati-chain \cite{Ames68}, i.e.
\begin{equation}
\ddot x+3x\dot x +x^3=0 \label{riceq}
\end{equation}
possesses many JLM and therefore Lagrangians. In particular  the following
time-independent Lagrangian\footnote{This Lagrangian has also been studied in
\cite{Cari05}.}  \begin{equation}Lagr=-\frac{1}{2(\dot
x+x^2)}\label{ricLagr}\end{equation} was shown to admit five Noether point
symmetries, i.e.
\begin{equation}
\Gamma_2-\Gamma_8, \quad \Gamma_3-\frac{2}{3}\Gamma_7, \quad \Gamma_4, \quad
\Gamma_5,\quad \Gamma_6
\end{equation}
among the following eight Lie symmetries admitted by equation (\ref{riceq}):
\begin{eqnarray}
\Gamma_1&=&t^3(t x-2)\partial_t-t(x t-2) (x^2 t^2+2-2 x t)\partial_x\nonumber\\
\Gamma_2&=&x t^3\partial_t-(x t-1) (x^2 t^2+4-2 x t)\partial_x\nonumber\\
\Gamma_3&=&x t^2\partial_t-x (x^2 t^2+2-2 x t)\partial_x\nonumber\\
\Gamma_4&=&x t\partial_t-x^2 (x t-1)\partial_x\nonumber\\
\Gamma_5&=&x\partial_t-x^3\partial_x\\
\Gamma_6&=&\partial_t\nonumber\\
\Gamma_7&=&t\partial_t-x\partial_x\nonumber\\
\Gamma_8&=&t^2\partial_t-2(x t-1)\partial_x.\nonumber
\end{eqnarray}
Lagrangian (\ref{ricLagr}) was obtained from the inverse of the determinant of
the matrix (\ref{Mnm}) with the two symmetries $\Gamma_5$ and $\Gamma_6$  and
the application of formula (\ref{lagrint}).

It interesting to show that equation (\ref{riceq}) is linked  to the following
 cubic-power dissipative equation
\begin{equation}
\frac{{\rm d}^2 \tilde x}{{\rm d}\tilde t^2}+\left(\frac{{\rm d} \tilde x}{{\rm
d}\tilde t}\right)^3=0\label{cubiceq}\end{equation}
 by means of the canonical form of the
two-dimensional Lie algebra generated by $\Gamma_5$ and $\Gamma_6$. In fact
this Lie algebra corresponds to Type I in Lie's classification of the real
two-dimensional Lie algebras in the plane \cite{Lie12}, \cite{Bianchi18},
namely it is abelian and transitive\footnote{In \cite{nuctam_3} a missprint led
to the erroneously statement that $\Gamma_5$ and $\Gamma_6$ generate an
intransitive Lie algebra.} and therefore its canonical coordinates are
\begin{equation}
\partial_{\tilde t}, \quad \quad \partial_{\tilde x}.
\end{equation}
Therefore the following identification
\begin{equation}
\Gamma_5=\partial_{\tilde t}, \quad \quad \Gamma_6=\partial_{\tilde x}
\end{equation}
yields the transformation
\begin{equation}
\tilde t= \frac{1}{2x^2}, \quad \quad \tilde x=\frac{tx-1}{x}
\end{equation}
that takes (\ref{riceq}) into (\ref{cubiceq}).

 Equation (\ref{cubiceq}) is
also linearizable and admits an eight-dimensional Lie symmetry algebra. Work is
in progress to quantize $n$-power dissipative equations with the method
illustrated here and to compare it to other proposed method, e.g.
\cite{Dieter07},  \cite{Dieter10}, \cite{Smith10}.\\

We now quantize the second-order Riccati equation (\ref{riceq}) by imposing
equation (\ref{sch2}) to admit the following five Lie symmetries
\begin{eqnarray}
\Gamma_2-\Gamma_8&\Rightarrow&\Lambda_1 =(xt-1) t^2\partial_t-(x t-1) (x^2
t^2-2-2 x t)\partial_x+
\lambda_1\partial_{\psi}, \nonumber\\[0.2cm]
\Gamma_3-\frac{2}{3}\Gamma_7&\Rightarrow&\Lambda_2 = \left(x
t-\frac{2}{3}\right)t\partial_t-x \left(x^2 t^2+\frac{4}{3}-2 x
t\right)\partial_x+ \lambda_2\partial_{\psi},
\nonumber \\[0.2cm]
\Gamma_4&\Rightarrow&\Lambda_3 =x t\partial_t-x^2 (x t-1)\partial_x+\lambda_3\partial_{\psi},\nonumber \\[0.2cm]
\Gamma_5&\Rightarrow&\Lambda_4 =x\partial_t-x^3\partial_x+\lambda_4\partial_{\psi},\nonumber \\[0.2cm]
\Gamma_6&\Rightarrow&\Lambda_5 =\partial_t+\lambda_5\partial_{\psi}.
\label{symschR}
\end{eqnarray}
where $\lambda_i=\lambda_i(t,x,\psi), (i=1,5)$ are functions of $t,x,\psi$ that
have to be determined.  We remind that equation (\ref{sch2}) also admits the
two symmetries (\ref{twosymb}).

Using the interactive REDUCE programs \cite{man}, we obtain that equation
(\ref{sch2}) becomes
\begin{equation}
4\psi_{tt}- 8x^2\psi_{tx}+4x^4\psi_{xx}+8x^3\psi_x-3x^2\psi=0 \label{schr}
\end{equation}
with
\begin{eqnarray}
\lambda_1=-\frac{(tx - 1)^3}{2x}\psi,\quad  \lambda_2=-\half (tx - 1)^2\psi,
\quad \lambda_3=-\half (tx - 1)x\psi \nonumber \\
\lambda_4=-\half x^2\psi, \quad \lambda_5=-\frac{tx - 1}{x}\psi.
 \end{eqnarray}
Equation (\ref{schr}) is parabolic  and therefore can be put into its normal
form since its characteristic coordinates are
\begin{equation}\varrho=t-\frac{1}{x},\quad x=x,\quad \psi=\phi(\varrho,x).\label{carcoor}
\end{equation} Thus
equation (\ref{schr}) transforms into
\begin{equation}
4x^2\phi_{xx}+8x\phi_x - 3\phi =0\end{equation} with solution:
\begin{equation}\phi(\varrho,x)=\beta_1(\varrho)x^{1/2}+\beta_2(\varrho)x^{-3/2},\end{equation}
where $\beta_1, \beta_2$ are arbitrary functions of $\varrho$. Is this the
correct quantization of equation (\ref{riceq})? It seems so but further insight
is needed especially from the experimentalists.

\section{Final remarks}
In \cite{gallipoli10} a method was proposed to overcome the deadlock of
nonlinear canonical transformations when quantizing with the known procedures
\cite{LGQM}. It consists of the following steps to be applied to a classical
Lagrangian equation\footnote{The physical Lagrangian turns out to admit the
highest number of Noether point symmetries.}:
\begin{itemize}
\item Find the Lie symmetries of the Lagrangian equation
$$\Upsilon=W_0(t,x)\partial_t+W_1(t,x)\partial_x$$
\item  Among the Lie symmetries find the Noether symmetries admitted
 by the given Lagrangian
$$\Gamma=V_0(t,x)\partial_t+V_1(t,x)\partial_{x}, \quad \quad \Gamma \subset \Upsilon$$
\item  Construct the Schr\"odinger equation admitting these
 Noether symmetries as Lie symmetries
$$2i\psi_t+f_{1}(x)\psi_{xx}+f_2(x)\psi_{x}+f_3(x)\psi=0$$
$$\Omega=V_0(t,x)\partial_t+V_1(t,x)\partial_{x}
+G(t,x,\psi)\partial_{\psi}$$
\item  Summarizing: quantize preserving the Noether symmetries
\end{itemize}
 In \cite{schgf} this method has
been  applied to classical Lagrangian systems. In particular it led to the
Schr\"odinger equation of a known completely integrable and solvable many-body
problem, the so-called `goldfish' \cite{calo01}. In the case of Lagrangian
systems the method consists of the following steps\footnote{The physical
Lagrangian turns out to admit the highest number of Noether point symmetries.}:
\begin{itemize}
\item  Find the Lie symmetries of the Lagrangian system
$$\Upsilon=W(t,\underline{x})\partial_t+\sum_{k=1}^{N}W_k(t,\underline{x})\partial_{x_k}$$
\item  Among the Lie symmetries find the Noether symmetries admitted
 by the given Lagrangian
$$\Gamma=V(t,\underline{x})\partial_t+\sum_{k=1}^{N}V_k(t,\underline{x})\partial_{x_k},\quad
\Gamma \subset \Upsilon$$
\item Construct the Schr\"odinger equation admitting these
 Noether symmetries as Lie symmetries
$$2i\psi_t+\sum_{k,j=1}^{N} f_{kj}(\underline{x})\psi_{x_jx_k}+
\sum_{k=1}^{N}h_k(\underline{x})\psi_{x_k}+f_3(\underline{x})\psi=0$$
$$\Omega=V(t,\underline{x})\partial_t+\sum_{k=1}^{N}V_k(t,\underline{x})\partial_{x_k}
+G(t,\underline{x},\psi)\partial_{\psi}$$
\item Summarizing: quantize preserving the Noether symmetries
\end{itemize}
In this paper we have proposed a method that takes classical equations into the
quantum realm by means of the Schr\"odinger equation regardless of the number
of Lagrangians that may classically exist. The method consists of the following
steps:
\begin{itemize}
\item  Find the Lie symmetries of the given equation
$$\Upsilon=W_0(t,x)\partial_t+W_1(t,x)\partial_x$$
\item Among many Lagrangians identify the one\footnote{They could be more than one
 as we have
shown in the present paper.}  that admits the highest number of Noether
symmetries among the Lie symmetries
$$\Gamma=V_0(t,x)\partial_t+V_1(t,x)\partial_{x}, \quad \quad \Gamma \subset \Upsilon$$
\item  Construct a linear partial differential equation
admitting these
 Noether symmetries as Lie symmetries\\
$f_{11}(t,x)\psi_{tt}+f_{12}(t,x)\psi_{tx}+f_{22}(t,x)\psi_{xx}+f_{1}(t,x)\psi_{t}
+f_2(t,x)\psi_{x}+f_0(t,x)\psi=0$
$$\Omega=V_0(t,x)\partial_t+V_1(t,x)\partial_{x}
+G(t,x,\psi)\partial_{\psi}$$
\item Since this  equation is parabolic, determine its
 canonical coordinates in order to transform it into its normal form, namely
 the Schr\"odinger equation
\item  Summarizing: quantize preserving the Noether symmetries.
\end{itemize}

\end{document}